# References of References: How Far is the Knowledge Ancestry

Chao Min
*Nanjing University*
Nanjing, China
mc@nju.edu.cn

Jiawei Xu
*Peking University*
Beijing, China
imxujiawei@pku.edu.cn

Tao Han
*Nanjing University*
Nanjing, China
hantaoer@foxmail.com

Yi Bu
*Peking University*
Beijing, China
buyi@pku.edu.cn

*Abstract*— Scientometrics studies have extended from direct citations to high-order citations, as simple citation count is found to tell only part of the story regarding scientific impact. This extension is deemed to be beneficial in scenarios like research evaluation, science history modelling, and information retrieval. In contrast to citations of citations (forward citation generations), references of references (backward citation generations) as another side of high-order citations, is relatively less explored. We adopt a series of metrics for measuring the unfolding of backward citations of a focal paper, tracing back to its knowledge ancestors generation by generation. Two sub-fields in Physics are subject to such analysis on a large-scale citation network. Preliminary results show that (1) most papers in our dataset can be traced to their knowledge ancestry; (2) the size distribution of backward citation generations presents a decreasing-and-then-increasing shape; and (3) citations more than one generation away are still relevant to the focal paper, from either a forward or backward perspective; yet, backward citation generations are higher in topic relevance to the paper of interest. Furthermore, the backward citation generations shed lights for literature recommendation, science evaluation, and sociology of science studies.

## I. INTRODUCTION

Simple citation count has long before been recognized for its inefficiency in estimating the impact of scientific works, as it may tend to favour applied research but undervalue basic science, bias towards a crowded field in comparison to a new field [1]. Among studies in search of measures on indirect scientific impact, Hu & Rousseau have proposed the notion of backward citation generations (references of references) and forward citation generations (citations of citations) [2]. Forward citation generations were investigated in virtue of a data structure called "citation cascade" in prior studies [3, 4]. Backward citation generations, however, are relatively underexplored. In this study, we examine the characteristics of backward citation generations on a large real citation network, aiming to find answers to the following questions:

- What characteristics do references of references exhibit along different reference generations? and
- What differences and similarities can we find between references of references and citations of citations?

## II. RELATED WORK

### A. The concepts of backward and forward citation generations

Citation is a relatively ambiguous concept: For a specific focal paper, one may not know a "citation" is whether another paper that it cites or another paper that cites it. To differentiate, usually the former is called a "reference" and the latter is called a "citation". The concepts of references and citations have been widely and multidimensionally analyzed by many big names including Henry Small in Library ad Information Science [5]. More recently, Hu, Rousseau & Chen adopted a consistent terminology by naming "reference" as backward citation and "citation" as forward citation. This terminology unifies the naming of citation in the timeline and raises the interesting concepts of backward and forward citation generations [2]. Hu et al. defined the focal paper as a zero-generation paper, thus obtaining backward citation generations by moving back in time and forward citation generations by moving forward in time [2]. Fragkiadaki & Evangelidis examined various definitions of citation generations in detail and discussed a number of indirect bibliometric indicators [6]. Both Hu et al. and Fragkiadaki et al. presented specific mathematical formulas for citation generations [2, 6]. Recently, forward citation generations seem to have received more attention than backward citation generations. Min, Sun & Ding for instance, observed citation structure from the angle of information cascade and quantified with two structural-wise metrics [4]. What they term as "citation cascades" is in essence a linked structure of forward citation generations. Chen also investigated this cascading citation expansion process which has potential to extend the coverage of relevant publications by citation links [7]. To model the evolution of follow-up works and their interdependence, Mohapatra et al. defined a data structure called Influence Dispersion Tree (IDT) [8]. Based on this data structure, they proposed a suite of metrics to measure the influence of a paper.

### B. The applications of backward and forward citation generations

Viewing citation data from the perspective of propagating generations has been applied in several studies. The first application one may think of is measuring the total influence in terms of a single paper, taking both direct and indirect citations into account. Rousseau did so from a backward citation perspective by using the Gozinto Theorem and claimed that this approach provides insight in the stream of ideas that guided an author to the results in a given paper [9]. della Briotta Parolo et

Supported by the National Science Foundation of China (Grant No. 71904081, No. 71874077) and Humanities and Social Sciences Program of the Ministry of Education (No. 19YJC870017).

XXX-X-XXXX-XXXX-X/XX/$XX.00 ©20XX IEEE

al. formalized the question of how much influence a seed publication *s* has on another publication *j* and called it *persistent influence*. They took into account all the citation passages between the two publications and calculated persistent influence in a large real citation network; they concluded that Nobel prize papers have higher ranks of persistent influence than direct citation counts [10]. Another stream of application is the investigation of citation chains. In their work, Frandsen & Nicolaisen documented a ripple effect of the Nobel Prize delivered through citation chains to mathematician Robert J. Aumann's scientific works as well as to those works' references. They showed that the award of the Nobel Prize affected not only the citations to Aumann's works but also the citations to those works' references [11]. The effect of citation chain reactions received follow-up attention from researchers [12, 13]. In addition, the concept of citation chain has also been applied in such tasks as search path counting [14] and main path analysis [15].

III. METHODS AND DATA

Adopting a similar approach to a data structure called "citation cascade" in Min, Chen & Yan et al. and Min, Sun & Ding [3, 4], we construct a data structure termed "reference cascade" from the perspective of backward citation. For each paper $i$, we define its reference cascade $RC_i$ as a directed graph that includes $i$, all references of $i$, second-order references (i.e., references of references) of $i$, third-order references, …, and $n$th-order references until no further references (called "ancestor papers"; they are the knowledge ancestry of $i$) can be identified; as for edges, $RC_i$ includes all citing relations within these publications. Thus, we know that the only node whose in-degree equals zero is the focal paper $i$, and that nodes whose out-degree equals zero represent its ancestor paper(s).

For each reference cascade, we define two metrics:

- **Depth**: A certain node's depth is the shortest path length from the focal paper $i$ to that node within the reference cascade. The maximum value of any node's depth is the depth of the reference cascade, which reflects how deep the reference cascade can reach.

- **Size**: Size is defined as the total number of nodes in a reference cascade. The size may dramatically increase as the depth of a reference cascade increases.

For each generation of a certain reference cascade, we define:

- **Width:** The number of nodes in a reference generation is called the generation's width. The maximum width of any generation is taken as the width of the reference cascade, which reflects how wide the reference cascade can extend.

- **Topic relevance:** Two papers' topic relevance is defined as the similarity of their topics, measured in the dataset here by the Jaccard similarity of their Physics and Astronomy Classification Scheme (PACS) codes. We then stipulate the topic relevance of a certain generation as the average topic relevance between the focal paper and all papers in this generation of the reference cascade. More details about Jaccard similarity and PACS codes can be found in the study [3].

In this study, we adopt the 2013 version American Physical Society (APS) dataset that covers ~450K physics publications and their ~6M citing relations. Publications after 1975 are labelled with one or more PACS codes to identify their sub-fields. As a pilot study, we particularly select publications whose PACS code starting with 2 (Nuclear physics; 19,516 papers) or 5 (Physics of gases, plasmas, and electric discharges; 5,113 papers) as focal papers.

IV. RESULTS

**Depth distribution.** Nearly 1500 (7.6% out of 19,516) and 600 (11.7% out of 5,113) papers respectively in PACS 2 and PACS 5 are recorded with no APS-indexed reference, as shown in Figure 1. We check the full texts of these papers and find that they are not necessarily very old papers who have few prior knowledge to cite, but quite often they have cited other sources of knowledge outside the APS literature. The remaining papers that have APS reference(s) more or less inherited knowledge from the dataset. The earliest ancestors in PACS 2 and PACS 5 date back to 61 and 55 generations away, respectively. In PACS 5, as the generation traces backward, a decreasing number of papers can be reached but there is a rise of papers between 35-45 generations. For PACS 2, the trend is a bit complicated as there are three noteworthy stages on the curve, firstly a decreasing stage between 1-18 generations, next an increasing-and-then-decreasing wave between 19-36 generations, and finally a repeating wave afterwards.

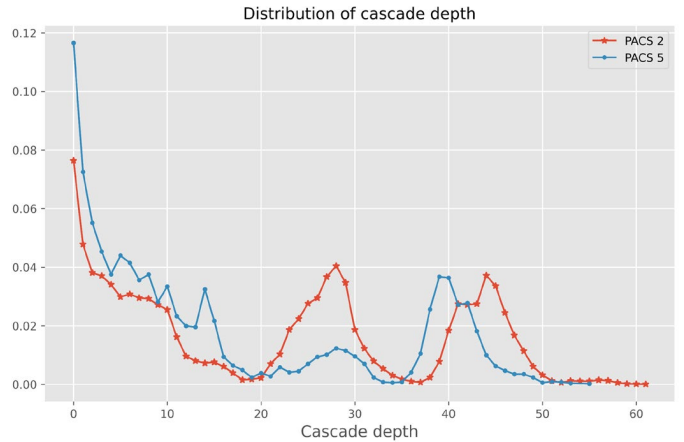

Fig. 1. Paper distribution of cascade depth.

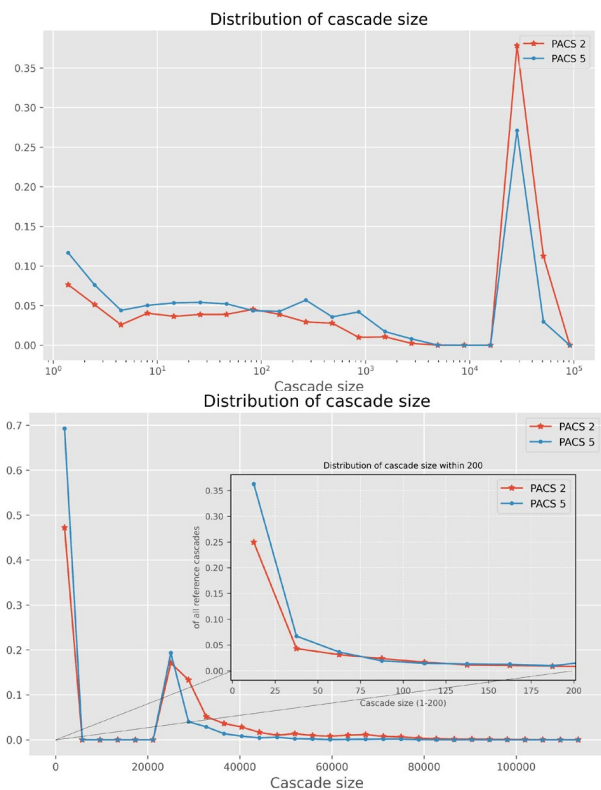

Fig. 2. Paper distribution of cascade size. Upper: log scale. Lower: normal scale and normal scale between 0 and 200.

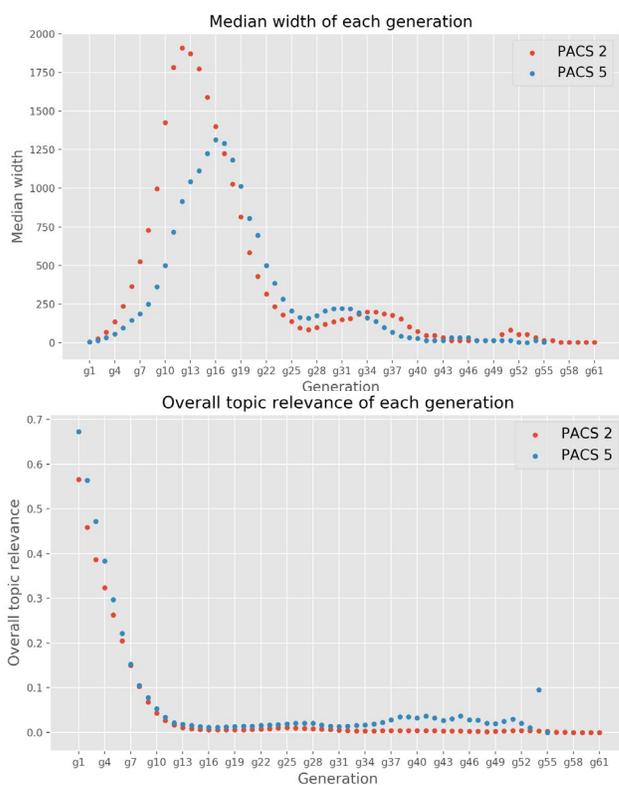

Fig. 3. Upper: Paper distribution of median width of each generation. Lower: Overall topic relevance across reference generations.

**Size distribution.** The size of reference cascades reaches a maximum of 115,384 in PACS 2, and 92,661 in PACS 5. From the upper sub-figure of Figure 2, we can see that the peak occurs in 20K-40K when using 20 bins with a logarithmic scale. Yet, with a normal scale (the lower sub-figure of Figure 2), it shows that the number of cascades less than 5K nodes accounts for more than 45% and 65% in PACS 2 and PACS 5, respectively. Aiming to understanding the distribution more comprehensively, we zoom in to investigate the distribution within 200 using a linear scale (see the lower inner sub-figure of Figure 2), and clearly observe a long-tail shape. An interesting finding in Figure 2 is that there seems to exist a vacuum zone between size of 5K-20K, where we observe almost no reference cascades in this range, but there do exist papers with either lower or higher sizes. This is in line with the observation on "citation cascades" in a prior study [3].

**Width distribution.** To explore how generation width (the number of papers in a generation) grows as references reach into prior knowledge space, we select the median value of generation width among all cascades (Figure 3 Upper). This gives a macro perspective on the general pattern of generation width growth. Both PACS 2 and PACS 5 show a similar trend on which the width increases from the first generation and then starts to decrease after reaching the peak. The peak appears at the $12^{th}$ generation for PACS 2 and four generations later for PACS 5. Interestingly, both width curves dip at the $27^{th}$ generation and then turn up for a bit afterwards.

**Topic relevance.** The overall topic relevance between a focal paper and a generation of prior papers is shown in Figure 3 (Lower). The first generation (that is, direct references), without doubt, presents a rather high relevance with the focal paper, nearly 0.7 for PACS 5 and 0.6 for PACS 2. Further generations, however, are not entirely irrelevant. Actually, the relevance is as high as 0.56 for PACS 5 and 0.46 for PACS 2 in the second generation. The third generation still has a relevance of 0.47 and 0.39 for PACS 5 and PACS 2, respectively. Furthermore, in both Physics sub-fields, references within six generations still have a relevance greater than 0.2. The relevance decreases sharply and stabilizes around the $16^{th}$ generation, after which the "ancestry" relationship seems rather weak.

## V. DISCUSSION

Studying reference cascades has far-reaching implications for science of science. First of all, reference cascades offer a feasible strategy of recommending related scientific literature for bibliographic databases and scholarly social network-based platforms (e.g., Mendeley). More specifically, those guiding the author(s) of the focal paper are explicitly reflected by its reference cascade, thus providing clearer clues on related publications. Rousseau mentioned that 2-4 generations of reference cascades might be sufficient for such a task [9], but the current paper extends his conclusions with more comprehensive empirical study.

Secondly, reference cascades may inspire more scientometric indicators for characterizing a focal publication, especially network-oriented ones, such as density, path-, degree-, and cluster-based measurements. These indicators depict a vivid picture on the process of knowledge integration (e.g., diversity of background knowledge). The current paper shows that topic relevance between the focal paper and within-six-generation references is greater than 0.2, offering empirical evidence of constraining the number of generations in a reference cascade for calculating meaningful measurements (yet, we should note that further empirical studies are still needed to dig into the effectiveness of this threshold). Meanwhile, we observe that if we only include ≤6 generations, the value of median width does not reach its peak (Figure 3 Upper), indicating the feasibility for computation. These might be quite beneficial for science policy makers and funding providers in practice.

Thirdly, this networked structure renders recorded documentations for digital library, sociology of science, and STS (science and technology studies). On the one hand, reference cascades are "fixed" once a focal paper is published, yet citation cascades may dynamically change over time. That being said, reference cascades offer an early signal for understanding the structure of a focal publication. On the other hand, since adding references is a norm in the current science system, most publications have reference cascades, as shown in Figure 1; but this is not the case for citation cascades. A previous descriptive study [16] has shown that there are more than 20% publications that are zero-cited (thus have no citation cascades), and this number might be underestimated due to the coverage of databases.

## VI. CONCLUSIONS AND FUTURE WORK

This paper reports the results of a research-in-progress study on backward citation generations. Different from many existing studies regarding forward citation generations, we trace backward into knowledge ancestry, or what we call references of references, of an individual paper. Two sets of papers from two sub-fields in Physics, as well as their backward citation generations, are extracted for analysis. Preliminary results show that: (1) Most publications in our dataset have reference cascades; (2) Many papers feature with a reference cascade of 0-5K (especially 0-100) or >20K nodes (i.e., size of cascade); yet we observe a vacuum zone between size of 5K-20K; (3) There are two peaks regarding median width of each generation, and the peaks occur in the $12^{th}$ and the $27^{th}$ generations, respectively; and (4) Publications within the sixth generations have obviously greater topic relevance with the focal publication (>0.2 using the Jaccard similarity) to the focal one. Since within several generations, backward citations are relevant to the paper of interest, a higher-generation approach would have potential over a direct-citation approach in seeking more citations that may be of one's interest. The proposal of reference cascades offers many applications for literature recommendation, science evaluation, and sociology of science studies.

However, further research is needed to more rigorously validate and extend the findings in this study using more comprehensive bibliographic datasets covering more disciplines; a discipline-wise comparison might help generalize current findings to more general domains. Besides, more easy-to-use scientometric indicators might be needed for high-generation citation structures. Meanwhile, future studies could compare reference cascades with citation cascades to paint an overview picture on the whole process of knowledge lifecycle, from its integration to use, from diffusion to death.

The current paper discusses three potentials of adopting the structure of reference cascades. Yet, details of how future researchers and science policy makers apply reference cascades to their specific studies/tasks are not deeply presented. We thus call for future research digging into applications of backward citations, including but not limited to the algorithms of literature recommendations, validations of networked indicators based on reference cascades, and theoretical underpinning of the dynamics of discipline paradigms, etc.

ACKNOWLEDGMENTS

The authors acknowledge the American Physical Society for publicly share their dataset for research.

REFERENCES

[1]  J. Margolis, "Citation Indexing and Evaluation of Scientific Papers," *Science*, vol. 155, no. 3767, pp. 1213–1219, Mar. 1967.

[2]  X. Hu, R. Rousseau, and J. Chen, "On the definition of forward and backward citation generations," *Journal of Informetrics*, vol. 5, no. 1, pp. 27–36, Jan. 2011.

[3]  C. Min, Q. Chen, E. Yan, Y. Bu, and J. Sun, "Citation cascade and the evolution of topic relevance," *J Assoc Inf Sci Technol*, vol. 72, no. 1, pp. 110–127, Jan. 2021.

[4]  C. Min, J. Sun, and Y. Ding, "Quantifying the evolution of citation cascades," *Proc. Assoc. Info. Sci. Tech.*, vol. 54, no. 1, pp. 761–763, Jan. 2017.

[5]  H. G. Small, "Cited Documents as Concept Symbols," *Soc Stud Sci*, vol. 8, no. 3, pp. 327–340, Aug. 1978.

[6]  E. Fragkiadaki and G. Evangelidis, "Review of the indirect citations paradigm: theory and practice of the assessment of papers, authors and journals," *Scientometrics*, vol. 99, no. 2, pp. 261–288, May 2014.

[7]  C. Chen, "Cascading Citation Expansion," *Journal of Information Science Theory and Practice*, vol. 6, no. 2, pp. 6–23, 2018.

[8]  D. Mohapatra, A. Maiti, S. Bhatia, and T. Chakraborty, "Go Wide, Go Deep: Quantifying the Impact of Scientific Papers Through Influence Dispersion Trees," in *2019 ACM/IEEE Joint Conference on Digital Libraries (JCDL)*, Champaign, IL, USA, Jun. 2019, pp. 305–314.

[9]  R. Rousseau, "The Gozinto theorem: Using citations to determine influences on a scientific publication," *Scientometrics*, vol. 11, no. 3–4, pp. 217–229, Mar. 1987.

[10] P. della Briotta Parolo, R. Kujala, K. Kaski, and M. Kivelä, "Tracking the cumulative knowledge spreading in a comprehensive citation network," *Phys. Rev. Research*, vol. 2, no. 1, p. 013181, Feb. 2020.

[11] T. F. Frandsen and J. Nicolaisen, "The ripple effect: Citation chain reactions of a nobel prize," *J Am Soc Inf Sci Tec*, vol. 64, no. 3, pp. 437–447, Mar. 2013.

[12] R. Farys and T. Wolbring, "Matched control groups for modeling events in citation data: An illustration of nobel prize effects in citation networks," *Journal of the Association for Information Science and Technology*, vol. 68, no. 9, pp. 2201–2210, Sep. 2017.

[13] T. F. Frandsen and J. Nicolaisen, "Rejoinder: Noble prize effects in citation networks," *Journal of the Association for Information Science and Technology*, vol. 68, no. 12, pp. 2844–2845, Dec. 2017.

[14] X. Jiang and H. Zhuge, "Forward search path count as an alternative indirect citation impact indicator," *Journal of Informetrics*, vol. 13, no. 4, p. 100977, Nov. 2019.

[15] X. Jiang, X. Zhu, and J. Chen, "Main path analysis on cyclic citation networks," *Journal of the Association for Information Science and Technology*, vol. 71, no. 5, pp. 578–595, May 2020.

[16] J. Beel, and B. Gipp. "Google Scholar's ranking algorithm: The impact of citation counts (An empirical study)," *2009 third international conference on research challenges in information science* (pp. 439-446). IEEE.